\documentclass[a4paper]{jpconf}
\usepackage{graphicx}
\begin{document}
\newcommand{\be}{\begin{equation}}
\newcommand{\ee}{  \end{equation}}
\newcommand{\ba}{\begin{eqnarray}}
\newcommand{\ea}{  \end{eqnarray}}
\newcommand{\ve}{\varepsilon}
\title{Nonequilibrium Green's functions in the study of heat transport of driven nanomechanical systems}

\author{Liliana Arrachea and Bruno Rizzo}

\address{Departamento de F'{\i}sica e IFIBA, Facultad de Ciencias Exactas y Naturales, Universidad de Buenos Aires. Pabell\'on I, Ciudad Universitaria (1428) Buenos Aires, Argentina}

\ead{lili@df.uba.ar}

\begin{abstract}
We review a recent theoretical development based  on non-equilibrium Green's function formalism to study heat transport in nanomechanical devices modeled by phononic systems
of coupled quantum oscillators driven by ac forces and connected to phononic reservoirs. We present the relevant equations to calculate the heat currents flowing along different regions of
the setup, as well as the power developed by the time-dependent forces. We also present different strategies to evaluate the Green's functions exactly or approximately within the weak
driving regime.  We finally discuss the different mechanisms in which the ac driving forces deliver the energy. We show that, besides generating heat, the forces may operate exchanging energy 
as a quantum engine.
\end{abstract}

%%%%%%%%%%%%%%%%%%%%%%%%%%%%%%%%%%%%%%%%%%%%%%%%%%%%%%%%%%%%%%%%%%%%%%%%%%%%%%%%%%%%%%%%%%%%%%%%%%%%%%%%%%%%%%%%%%%%%
\section{Introduction}
The Green's function formalism provided the theoretical framework for many studies of electron transport in mesoscopic devices. Since the pioneer work by Caroli and coworkers, \cite{car} the
celebrated generalized Landauer-B\"uttiker formula by Meir and Wingreen \cite{mei-wi} and the generalization to setups driven by the time-dependent voltages and fields of Refs. \cite{past,meiwijau}, 
several interesting applications can be found in the literature. Details of these historical developments 
and some important aspects of the subsequent  progress have been reviewed in papers of this conference series \cite{jaurev1,jaurev2}.

 In the last years, transport experiments carried out in the presence of
ac driving fields and the study of the related pumping phenomena \cite{acpump1,acpump2,acpump3,acpump4,acpump5,acpump6,acpump7} motivated the development of strategies to formulate and efficiently solve the corresponding Dyson's equations in 
theoretical models for these setups. A rigorous formulation was presented in Refs. \cite{liliring} for the problem of a mesoscopic ring threaded by a time-dependent magnetic flux. Such treatment was afterwards adapted to study particle transport in models of quantum pumps consisting in quantum dots where local ac voltages are applied, with \cite{lilipumpk} and without many-body interactions \cite{lilipumps}. In the latter case, an explicit relation between the Green's function and the Floquet scattering matrix was  presented in Ref. \cite{lilimos}.

Microscopic models and  the framework provided by non-equilibrium Green's functions also offer the adequate basis for the  study of mechanisms for heat generation, energy transport and quantum refrigeration by electrons in mesoscopic structures under ac driving, \cite{liliheat} as well as the possibility of characterizing these processes by defining effective temperatures \cite{fdt}. In the context of  phononic systems, J-S Wang and coworkers proposed recently in Ref. \cite{wang} a treatment analogous to that of Ref. \cite{car} to describe thermal transport induced by a temperature gradient under stationary conditions.
 The present contribution is devoted to review a recent related development, which focuses on the study of heat transport by phonons under ac driving \cite{liliphon}. One of the motivations is to
 investigate
  cooling mechanisms in these systems. This is because
  refrigeration by means of manipulating phonons, rather than electrons, has the appealing feature of being suitable for insulating as well as metallic systems. On the other hand,
 cooling nanomechanical systems is a very  active field of research \cite{coolnan0,coolnan1,coolnan2,coolnan3,coolnan4,coolnan5,coolnan6} and it is interesting to explore alternative mechanisms. 
 
 In a recent work,\cite{us} it was shown that when a barrier (or
perturbation) moves along a cavity connecting
two gases of acoustic phonons, a sizable amount of heat can be
transferred from the colder to the hotter gas. When the process is repeated as a cycle, the result is refrigeration until the colder reservoir reaches a minimum temperature determined by geometrical parameters of the device and the velocity at which the barrier travels through the cavity. That mechanism was afterward analyzed in the framework of a microscopic model which was
solved by means of a nonequilibrium Green's functions treatment analogous to the one proposed to study heat transport in ac driven electron systems. In the subsequent sections we briefly review that theoretical treatment. We also present some approximations to  
 study the problem in the limit of weak driving. Finally, we show that, besides refrigeration, another interesting mechanism suggested to take place in quantum pumps \cite{liliheat} and quantum capacitors 
 \cite{mobu}, can be also realized in phononic systems. Namely, the exchange of energy between the driving fields.

%%%%%%%%%%%%%%%%%%%%%%%%%%%%%%%%%%%%%%%%%%%%%%%%%%%%%%%%%%%%%%%%%%%%%%%%%%%%%%%%%%%%%%%%%%%%%%%%%%%%%%%%%%%%%%%%%%%%%
%%%%%%%%%%%%%%%%%%%%%%%%%%%%%%%%%%%%%%%%%%%%%%%%%%%%%%%%%%%%%%%%%%%%%%%%%%%%%%%%%%%%%%%%%%%%%%%%%%%%%%%%%%%%%%%%%%%%%
\section{Background}
\label{sec:back}
%
%%%%%%%%%%%%%%%%%%%%%%%%%%%%%%%%%%%%%%%%%%%%%%%%%%%%%%%%%%%%%%%%%%%%%%%%%%%%%%%%%%%%%%%%%%%%%%%%%%%%%%%%%%%%%%%%%%%%%
\subsection{Microscopic Model}\label{sec:theory}
%\subsection{Model} 
%

The standard microscopic model for acoustic phonons consists of masses
coupled by springs. The one-dimensional (1d) version of this model became popular in the context  of heat  transport since the pioneer work by Fermi, Pasta, and Ulam \cite{fpu} 
and several subsequent studies devoted to understand the violation of the Fourier law in these systems \cite{wang,modpho1,modpho2,modpho3, revpho1,revpho2}. Fourier law implies a linear drop of the local temperature along a system connected
between reservoirs at different temperatures. It is the counterpart of Ohm law in the context of heat transport, and it is nowadays rather natural to accept that it should not be
expected to hold through a finite chain of harmonic oscillators, along which heat propagates coherently. In fact, a similar effect is known to take place in the behavior of the local voltage along a mesoscopic structure 
placed between two reservoirs at different chemical potentials \cite{cont-res1,cont-res2,cont-res3}. In these systems, the voltage drop takes place at the contacts and the concept of ''contact resistance'' was introduced to characterize this feature, while it vanishes along the structure through which electrons propagate coherently.
It is remarkable that the studies on the origin of the lack of the Fourier law in phononic systems and the introduction of the concept of the contact resistance in electron systems  followed parallel routes
and the connection between these two phenomena has not been mentioned even in review articles \cite{revpho1,revpho2}.
 More recently, heat pumping has
been analyzed through a two-level system driven by an ac field coupled
to phononic baths at different temperatures,~\cite{segal} between
semi-infinite harmonic chains with a time-dependent
coupling,~\cite{li} and in anharmonic molecular junctions in contact
to phononic baths \cite{moljun}.

%\begin{verbatim}
\begin{figure}[h]
\begin{center}
\includegraphics[width=35pc]{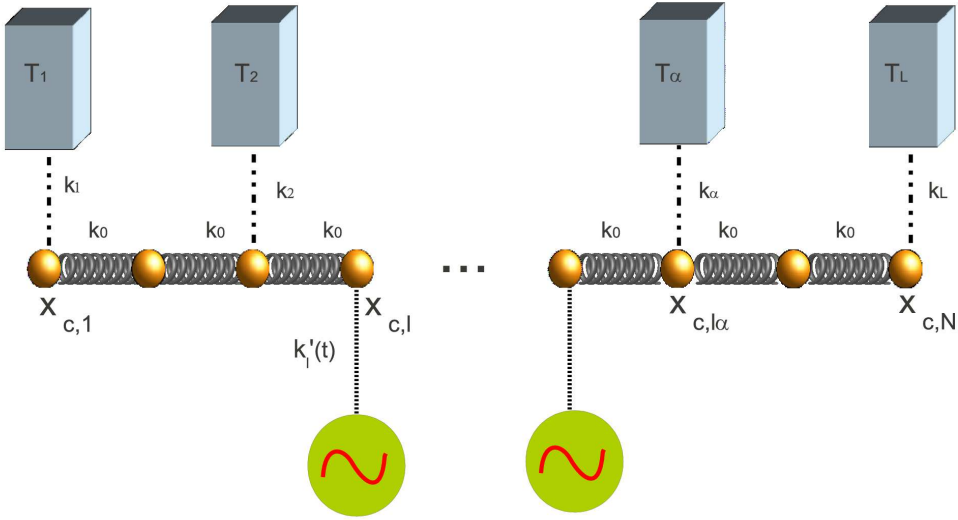}
\end{center}
\caption{\label{fig1}Sketch of the setup containing a central chain of acoustic phonons represented by a 1d lattice of atoms with equal masses connected by springs. This system is in contact to
several reservoirs at different temperatures and affected by driving forces acting locally at the different positions of the central chain.}
\end{figure}
%\end{verbatim}

We consider the one-dimensional (1d) system sketched in Fig. \ref{fig1},
with a central finite chain of
$N$ molecules with identical masses $m_c$ coupled by springs of constant
$k_0$. This system is connected, at certain sites $l_{\alpha}$, to 
semi-infinite chains of masses $m_{\alpha}$ coupled by
springs of constant $k_{\alpha}$. These chains play
the role of reservoirs, which we assume being kept at temperatures
$T_{\alpha}$. 
\begin{equation}
\label{H}
H=\sum_{\alpha=1}^{L} H_{\alpha}+H_{c}(t)+H_{\rm cont}.
\end{equation}
The first term $H_{\alpha}$ represents the reservoirs; the second
$H_c(t)$ describes the central chain. We assume that external time-dependent perturbations are applied at this system. The Hamiltonian reads
\begin{eqnarray}
\label{hcen}
H_{c}(t) & = & \sum_{l=1}^{N} \frac{p_{c,l}^2}{2 m_{c}} +
\sum_{l=1}^{N-1}\frac{k_0}{2} (x_{c,l}- x_{c,l+1} )^2 
+\ \sum_{l=1}^{N} \frac{k^{\prime}_{l}(t)}{2} \; x_{c,l}^2,
\end{eqnarray}
where the last term of this expression describes the time dependent perturbation, which is represented by  time-dependent elastic forces
at different positions of the central chain. 
It is useful to define two characteristic frequencies in the problem: the
 frequency $\omega_c = \sqrt{k_0/m_c}$, and the driving frequency
$\Omega_0$.(We shall work on units such that $\hbar= k_B=1$). The contact between the central chain and the reservoirs is described by the Hamiltonian
\be 
\label{hcon}
H_{\rm cont} =\sum_{\alpha=1}^{L} \frac{k_{\alpha}}{2} (x_{\alpha,1} - x_{c,l_{\alpha}})^2.
\ee
Notice that the site  $j=1$  of each reservoir couples to  the
site $l_{\alpha}$ of the central region.
The reservoir Hamiltonians are given by
\begin{equation}
\label{half}
H_{\alpha} = \sum_{j=1}^{N_{\alpha}}\left[ \frac{p_{\alpha,j}^2}{2
    m_{\alpha}} + \frac{k_{\alpha}}{2} (x_{\alpha,j} -
  x_{\alpha,j+1})^2 \right].
\end{equation}
In the limit of semi-infinite chains, $N_{\alpha} \rightarrow \infty$,
it is convenient to express the degrees of freedom of the
reservoirs in terms of normal modes. For open boundary
conditions, this corresponds to performing the following
transformation,
\begin{equation}
x_{\alpha, l} = \sqrt{ \frac{2}{ N_{\alpha}+1 } } 
\sum_{n=0}^{N_{\alpha}} \sin (q^{\alpha}_n l) \; x_{\alpha, n}
\end{equation}
and
\begin{equation}
p_{\alpha, l} = \sqrt{ \frac{2}{N_{\alpha}+1}} 
\sum_{n=0}^{N_{\alpha}} \sin (q^{\alpha}_n l) \; p_{\alpha, n},
\end{equation}
where
\be 
\label{qal} 
q^{\alpha}_n = \frac{n\pi}
{N_{\alpha}+1},\;\;\;\;n=0,\ldots,N_{\alpha}.
\ee
The corresponding Hamiltonians transform into
\be 
H_{\alpha} = \sum_n \left\{\frac{p_{\alpha,n}^2}{2 m_{\alpha}} +
\frac{k_{\alpha}}{2} \left[ 1- \cos(q^{\alpha}_n) \right]
x_{\alpha,n}^2 \right\}. 
\ee
%%%%%%%%%%%%%%%%%%%%%%%%%%%%%%%%%%%%%%%%%%%%%%%%%%%%%%%%%%%%%%%%%%%%%%%%%%%%%%%%%%%%%%%%%%%%%%%%%%%%%%%%%%%%%%%%%%%%%%
\subsection{Energy balance}
\label{sec:energy}
\begin{figure}[h]
\begin{center}
\includegraphics[width=35pc]{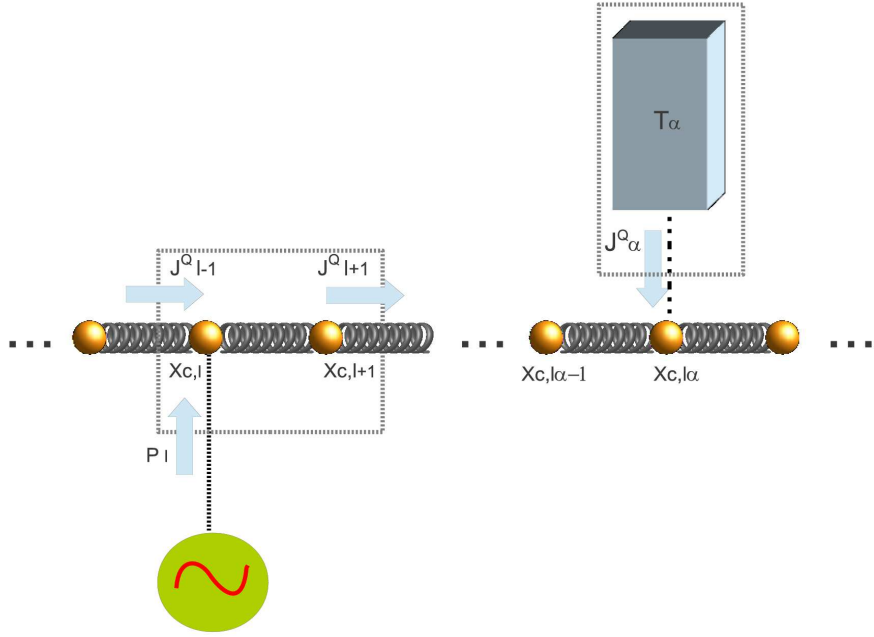}
\end{center}
\caption{\label{fig2}Sketch of the setup indicating the construction employed to derive the expressions for the heat currents as well as the poser developed by the external forces.}
\end{figure}

Following a procedure similar to that used in
Ref.~\cite{liliheat}, we consider the evolution in time of the
energy density stored at a given elementary volume and write the corresponding conservation equation. 
Since we are treating a 1d lattice, and our Hamiltonians involve only nearest-neighbors terms, the minimum volume 
that we consider is one that encloses nearest-neighbors sites, like the left box of Fig. \ref{fig2}. If we denote by $dE_{l,l+1}/dt$ the total energy flow exiting 
and entering a volume of the central chain without connection to reservoirs, the equation for the conservation of the energy for reads:
\begin{equation}
\frac{dE_{l,l+1}}{dt}=J_{l+1}^{Q}(t)-J_{l}^{Q}(t)+P_{l}(t)+P_{l+1}(t),
\end{equation}
where $J_{l}^{Q}(t)$ denotes the incoming energy current from site $l-1$ towards the site $l$ and $J_{l+1}^{Q}(t)$ the outgoing flow from site $l+1$ to $l+2$. We denote with a positive sign the flows pointing from left to right. These quantities are evaluated from
\be
\frac{dE_{l,l+1}}{dt}=-i\langle [H,H_{l,l}+H_{l+1,l+1}+H_{l,l+1}]  \rangle + \frac{d[H_{l}+H_{l+1}](t))}{dt},
\ee
where $H_{l,l^{\prime}}$ are the matrix elements of $H_c(t)$ on the sites $l,l^{\prime}$. The explicit calculation cast
\be
J_{l}^{Q}=\frac{k_0}{m_c}\langle p_{c,l}x_{c,l-1} \rangle,
\ee
for the current and
\begin{equation}
\label{powla}
P_{l}(t) = \frac{\partial k^{\prime}_{l}(t)} {\partial t} \,
 \,\langle x_{c,l}^2 \rangle,
\end{equation}
for the power developed by the force applied at the site $l$.

%If we enclose contacts bonds, we find
%
%\begin{eqnarray}
%\label{cer}
%&& \frac{dE_{i_{\alpha},i_{\alpha}+1}}{dt} = J_{i+1}^{E}(t)-J_{i}^{E}(t)-J^E_{\alpha+1}(t)-J^E_{\alpha}(t)+P_{i_\alpha}(t)+ %P_{i_{\alpha}+1}(t),
%\end{eqnarray}
% 
%where the $J^E_{\alpha}(t)$ term is the energy current flowing from the connecting
%site $i_\alpha$ of the central chain into the $\alpha$ reservoir. 

%The explicit expressions for the flows and powers are derive by recourse of the Eherenfest theorem:
%\begin{eqnarray}
%\frac{dE_{i,i+1}}{dt}&=&-\frac{i}{\hbar}\left\langle\left[H,\frac{p_{c,i}^{2}}{2m_c}+\frac{p_{c,i+1}^{2}}{2m_c}+\frac{k_0}
%{2}(x_{c,i}-x_{c,i+1})^{2}+\frac{k_{i}^{\prime}}{2} x_{c,i}^{2}+\frac{k_{i+1}^{\prime}}{2} x_{c,i+1}^{2} \right]\right\rangle\\
%&& + \frac{\partial k_{i}^{\prime}}{\partial t} \langle x_{i}^{2}\rangle+\frac{\partial k_{i+1}^{\prime}}{\partial t} \langle %x_{i+1}^{2}\rangle.
%\end{eqnarray}
%
%%%%%%%%%%%%%%%%%%%%%%%%%%%%%%%%%%%%%%%%%%%%%%%%%%%%%%%%%%%%%%%%%%%%%%%%%%%%%%%%%%%%%%%%%%%%%%%%%%%%%
\subsubsection{Heat flow into reservoirs}
We now focus on the right box of the sketch of Fig. \ref{fig2}. The variation in time of the energy stored in the reservoir $\alpha$ is
\begin{eqnarray}
\frac{dE_{\alpha}(t)}{dt}&=&-i \langle \left[H,H_{\alpha}\right] \rangle  = J_{\alpha}^{Q}(t),
%\left\langle \left[H,\sum_{n}\frac{p_{\alpha,n}^2}{2 m_{\alpha}} +
%\frac{k_{\alpha}}{2} ( 1- \cos(q^{\alpha}_n))x_{\alpha,n}^2\right]\right\rangle + \frac{\partial} {\partial t}
%k^{\prime}_{i_{\alpha}}(t) \,\langle x_{c,i_{\alpha}}^2  \rangle  \\
%&=& J_{\alpha}^{E}(t)+P_{i_{\alpha}}(t).
\end{eqnarray}
where we assume that no driving force is acting at the contact site of the reservoir. The current $J_{\alpha}^{Q}(t)$ flows along the contact between the site
 $l_{\alpha}$ of the central chain and the reservoir $\alpha$ and it is assumed to be positive when it enters the reservoir.
Since we are dealing with phonons, this energy flow is a pure heat flow and its explicit expression reads
\begin{equation}
\label{curalft}
J^Q_{\alpha}(t)= \sum_n \frac{k_{\alpha}}{m_c}\sqrt{ \frac{2}{ N_{\alpha}+1 } } \sin
(q^{\alpha}_n) \langle x_{\alpha,n}\hspace{0.1cm}p_{c,l_{\alpha}} \rangle .
\end{equation}
Conservation of energy implies that the {\em total} average of the power invested by
the external fields is dissipated into the reservoirs at a rate
\be \label{cons} \sum_{\alpha=1}^{L}\overline{J}^Q_\alpha=  \sum_{l=1}^{N}
\overline{P}_{l}, \ee
where
\be \overline{P}_{l}= \frac{1}{\tau}\int_0^{\tau} dt\, P_l(t), \ee
with $\tau=2\pi/\Omega_0$, is the power developed by the external fields at the site $l$ averaged
over one period, while
\be \label{dcj} \overline{J}^Q_{\alpha}= \frac{1}{\tau}\int_0^{\tau}
dt\, J^Q_{\alpha}(t), \ee
is the dc component of the heat current at the reservoir ${\alpha}$.
Notice that a net amount of work has to be invested in order to pump
heat from one reservoir to the other. In the next sections we show that carrying out an analysis similar
to that developed in Ref. \cite{liliheat} leads to the
conclusion that the rate at which the total work done by the external
fields is dissipated as heat flowing into the reservoirs is
proportional to $\Omega_0^2$.
%
%In the case of heat transport induced by a temperature gradient
%between the reservoirs, and in the absence of time-dependent
%perturbations, the heat current (\ref{dcj}) can be in general
%expressed as
%
%\be \label{estalph}
%J^{(0)}_{\alpha}=\sum_{\beta=1}^{L}\int_{-\infty}^{+\infty} \frac{d
%  \omega}{2 \pi} \omega [n_{\beta}(\omega)-n_{\alpha}(\omega)] {\cal
%  T}(\omega), \ee
%
%where ${\cal T}(\omega)$ is the thermal transmission function of the
%structure, while $n_{\alpha}(\omega)$ is the Bose-Einstein
%distribution function, which depends of the temperature of the
%reservoir $\alpha$.
%
%We evaluate these physical quantities by employing the
%non-equilibrium Green function formalism. Before describing the
%technical details of the calculations, we first present the results
%that demonstrate the refrigeration capabilities of the phonon pump.
%
%%%%%%%%%%%%%%%%%%%%%%%%%%%%%%%%%%%%%%%%%%%%%%%%%%%%%%%%%%%%%%%%%%%%%%%%%%%%%%%%%%%%%%%%%%%%%%%%%%%%%%%%%%%%%%%%%%%%%
%%%%%%%%%%%%%%%%%%%%%%%%%%%%%%%%%%%%%%%%%%%%%%%%%%%%%%%%%%%%%%%%%%%%%%%%%%%%%%%%%%%%%%%%%%%%%%%%%%%%%%%%%%%%%%%%%%%%%
\section{Non-equilibrium Green's function approach}
\subsection{Relevant Green's functions}
\label{sec:green}
%
%
%
%%%%%%%%%%%%%%%%%%%%%%%%%%%%%%%%%%%%%%%%%%%%%%%%%%%%%%%%%%%%%%%%%%%%%%%%%%%%%%%%%%%%%%%%%%%%%%%%%%%%%%%%%%%%%%%%%%%%%%%
The theory of non-equilibrium Green's functions, which was developed independently by  L. P. Kadanoff and G. Baym \cite{kadanoff}, Schwinger \cite{schwinger} and Keldysh \cite{keldysh} has been
employed many times in the context of electron transport, as mentioned in the introduction. Details can be found in that literature. In the present case, the general procedure is completely analogous.
 The relevant Green's functions are the bigger and lesser functions
\begin{eqnarray}
&& D^{>}(1,1^{\prime})=i\langle X_1(t_1) X_{1^{\prime}}(t_{1^{\prime}}) \rangle, \nonumber \\
& & D^{<}(1,1^{\prime})=i\langle X_{1^{\prime}}(t_{1^{\prime}}) X_{1}(t_1) \rangle,
\end{eqnarray}
where  $X(1)= \sqrt{m_1}x_{1}(t_1)$ is a phononic operator and $1,1^{\prime}$ is a schematic notation that labels the spacial and time coordinates.
The corresponding retarded and advanced Green's functions read
\begin{eqnarray}
\label{retarded}
&& D^{R}(1,1^{\prime})=\theta(t_1-t_{1^{\prime}})[D^{>}(1,1^{\prime})-D^{<}(1,1^{\prime})] \\
&& D^{A}(1,1^{\prime})=\theta(t_{1^{\prime}}-t_1)[D^{<}(1,1^{\prime})-D^{>}(1,1^{\prime})],
\end{eqnarray}
where the four Green functions satisfy the following relations
 $D^{A}(1,1^{\prime})=[D^{R}(1^{\prime},1)]^{*}$, $D^{<}(1,1^{\prime})=-[D^{>}(1^{\prime},1)]^{*}$ and 
$D^{R}(1,1^{\prime})-D^{A}(1,1^{\prime})=D^{>}(1,1^{\prime})-D^{<}(1,1^{\prime})$.
\subsection{Dyson's equation}
We now turn to derive the Dyson's equations to evaluate the Green's functions in our problem. The operators $x$ and $p$ satisfy canonical commutation relations:
\begin{eqnarray}
&& [x_{l},x_{l^{\prime}}]=0,\;\;\;\;\; [x_{l},p_{l^{\prime}}]=i\delta_{l,l^{\prime}}.
\end{eqnarray} 
For coordinates $l, l^{\prime}$ within the central chain, the lesser and greater Green's functions read
\begin{eqnarray}
&& D_{l,l^{\prime}}^{<}(t,t^{\prime})=i m_c \langle x_{c,l^{\prime}}(t^{\prime}) x_{c,l}(t), \rangle \\
&& D_{l,l^{\prime}}^{>}(t,t^{\prime})=i m_c \langle x_{c,l}(t) x_{c,l^{\prime}}(t^{\prime})  \rangle,
\end{eqnarray}
while the retarded Green's function is
\be
D_{l,l^{\prime}}^{R}(t,t^{\prime})=-i\theta(t-t^{\prime})m_c \langle[x_{c,l}(t),x_{c,l^{\prime}}(t^{\prime})] \rangle.
\ee
Now we proceed to the calculation of the equation of motion of the retarded Green's function. To this end we must take the second derivative respect to $t^{\prime}$,  and use the Eherenfest's theorem.
The result is
%\begin{eqnarray}
%\label{first}
%-i\hbar \frac{\partial D^{R}_{i,i^{\prime}}(t,t^{\prime})}{\partial t^{\prime}}=-\frac{i\hbar}{m_c}\theta(t-t^{\prime})\langle x_{i}%(t)p_{i^\prime}(t^{\prime}) \rangle.
%\end{eqnarray}
%If one more time to Eq.(\ref{first}), we obtain the following result for any site $i^{\prime}\neq 1$ or $N$:
%
\begin{eqnarray}
\label{dyson}
&&-\left[\frac{\partial ^{2}}{\partial t^{\prime ^{2}}}+2\frac{k_0}{m_c}+\frac{k_{l^{\prime}}^{\prime}(t)}{m_c}\right]D^{R}_{l,l^{\prime}}(t,t^{\prime}) 
+\frac{k_0}{m_c}\left[D^{R}_{l,l^{\prime}-1}(t,t^{\prime})+ D^{R}_{l,l^{\prime}+1}(t,t^{\prime})\right]
\\
&&=\delta_{l,l^{\prime}}\delta(t-t^{\prime})+\frac{k_{\alpha}}{m_c} \left[ D^{R}_{l,l_{\alpha}}(t,t^{\prime})\delta_{l^{\prime},l_{\alpha}}-D^{R}_{l,\alpha_{1}}(t,t^{\prime}) \delta_{l^{\prime},l_{\alpha}}\right],
\end{eqnarray}
where $l_{\alpha}$ refers to the site of the central chain connected to reservoir $\alpha$ and $\alpha_{1}$ the first site of the same reservoir.  In turn, the integral representation for the Dyson's equation  corresponding to the function
$D^{R}_{l,\alpha_{1}}$ reads
\be
\label{dyson2}
D_{l,\alpha_{1}}^{R}(t,t^{\prime})=-\frac{k_{\alpha}}{m_{\alpha}}\sum_{n=0}^{N_{\alpha}}\frac{2}{N_{\alpha}+1}\sin^{2}{(q_{n}^{\alpha})}\int_{-\infty}^{+\infty} dt_{1} D^{R}_{i,l_{\alpha}}(t,t_{1}) d^{R}_{q_{n}^{\alpha}}(t_{1},t^{\prime}),
\ee
where
\be d^R_{q^{\alpha}_n}(\omega) = \frac{1}{2 E_{\alpha,n}} \left[
  \frac{1}{\omega+ i \eta - E_{\alpha,n} }- \frac{1}{\omega+ i \eta +
    E_{\alpha,n}} \right], \ee
with $\eta=0^+$ stands for the equilibrium Green's function within the uncoupled reservoir $\alpha$. If we replace the Eq.(\ref{dyson2}) into Eq.(\ref{dyson}), we can define the self energy as
\be \label{sigal}
\Sigma_{\alpha}^{R}(t,t^{\prime})= \frac{k_{\alpha}^{2}}{m_{c} m_{\alpha}} \sum_{n=0}^{N_{\alpha}}\frac{2}{N_{\alpha}+1}\sin^{2}{(q_{n}^{\alpha})} d^{R}_{q_{n}^{\alpha}}(t,t^{\prime}),
\ee
(a more explicit expression is presented in  \ref{apa}),  and Eq.(\ref{dyson}) can be written in the simpler matrix form
\begin{eqnarray}
\label{dyson3}
-\partial_{t^{\prime}}^{2} \hat{D}^{R}(t,t^{\prime})+\hat{D}^{R}(t,t^{\prime})\hat{M}(t^{\prime})-\int dt_1 \hat{D}^{R}(t,t_1)\hat{\Sigma}^{R}(t_1,t^{\prime})=
\delta(t-t^{\prime}),
\end{eqnarray}
with $\hat{M}(t)=\hat{M}^{(0)}+\hat{M}^{(1)}(t)$, where
\begin{equation}
\label{m0}
 M^{(0)}_{l,l^{\prime}} = \left\{ \begin{array}{cc} \frac{k_0}{m_c} ( 2
   \delta_{l,l^{\prime}} - \delta_{i^{\prime},i \pm 1})+\frac{k_{\alpha}}{m_c}\delta_{l^{\prime},l_{\alpha}}\delta_{l,l^{\prime}}, & 1<l<N, \\ (
   \frac{k_0}{m_c}+ \frac{k_1}{m_c}) \delta_{l,l^{\prime}} - \frac{k_0}{m_c}
   \delta_{l^{\prime},l + 1}, & i=1, \\ ( \frac{k_0}{m_c}+ \frac{k_N}{m_c})
   \delta_{l,l^{\prime}} - \frac{k_0}{m_c} \delta_{l^{\prime},l - 1}, &
   l=N, \end{array} \right.
\end{equation}
\be
\label{mt}
M_{l,l^{\prime}}^{(1)}(t)=\frac{k_{l}^{\prime}(t)}{m_c}\delta_{l,l^{\prime}},
\ee
and $\Sigma_{l,l^{\prime}}^{R}(t,t^{\prime})=\sum_{\alpha=1}^{L}\delta_{l,l^{\prime}}\delta_{l,l_{\alpha}} \Sigma_{\alpha}^{R}(t,t^{\prime})$.

On the another hand, the Dyson's equation for the lesser function is obtained by recourse to Langreth's rules
%online
\ba
\label{dyless}
D^<_{l,l^{\prime}}(t,t^{\prime}) & = & \sum_{\alpha} \int dt_1 \int
dt_2\, D^R_{l,l_{\alpha}}(t,t_1)\, \Sigma_{\alpha}^<(t_1-t_2) D^A_{l_{\alpha},l^{\prime} }(t_2,t^{\prime}),
\ea
and \ba
\label{dyless1}
 D^<_{l,q^{\alpha}_n}(t,t^{\prime}) & = &
 -\frac{k_{\alpha}}{m_{\alpha}}\sum_{n}\sqrt{\frac{2}{N_{\alpha}+1}}\sin(q_{n}^{\alpha}) \int dt_1 [
   D^R_{l,l_{\alpha}}(t,t_1)\, d^<_{q^{\alpha}_n}(t_1-t^{\prime})
   \nonumber \\ & & +\ D^<_{l,l_{\alpha} }(t,t_1)\,
   d^A_{q^{\alpha}_n}(t_1-t^{\prime})], \ea
where Eq. (\ref{dyless}) corresponds to coordinates along the central
chain while Eq. (\ref{dyless1}) corresponds to one coordinate along
the chain and the other on the reservoir.  The advanced Green's function of the uncoupled reservoir is $d^{A}_{q_{n}^{\alpha}}(t_{1},t^{\prime})=[d^{R}_{q_{n}^{\alpha}}(t^{\prime},t_{1})]^*$,
while the corresponding lesser function reads%
\be d^<_{q^{\alpha}_n}(\omega) = \frac{i \pi n_{\alpha}(\omega)}{
  E_{\alpha,n}} \left[ \delta(\omega- E_{\alpha,n} )-\delta(\omega+
  E_{\alpha,n}) \right], \ee
with $n_{\alpha}(\omega)= 1/(e^{\omega/T_{\alpha}} - 1)$ being the
Bose-Einstein distribution function, which depends on the temperature
$T_{\alpha}$ of the reservoir $\alpha$. In terms of the spectral function defined in \ref{apa}, the Fourier transform of the lesser Green's function reads
\be \Sigma_{\alpha}^<(\omega) = i n_{\alpha}(\omega)
\Gamma_{\alpha}(\omega).\ee
%
%%%%%%%%%%%%%%%%%%%%%%%%%%%%%%%%%%%%%%%%%%%%%%%%%%%%%%%%%%%%%%%%%%%%%%%%%%%%%%%%%%%%%%%%%%%%%%%%%%
\subsection{Solution of Dyson's equation}
To obtain the retarded function of the central region, $\hat{D}^{R}$, one has to solve the differential equations (\ref{dyson3}). For the case of terms that depend harmonically on time,
it is convenient to follow the strategy proposed in Refs.  \cite{liliring,lilipumps} for electron systems and recently adapted to phonons in Ref. \cite{liliphon}. We summarize the main steps in 
the present section. It is convenient to perform the Fourier transform with respect to the "delayed time" in the retarded Green's function
\be
\hat{D}^{R}(t,\omega)=\int_{-\infty}^{+\infty}dt^{\prime}e^{i(\omega +i0^{+})(t-t^{\prime})}\hat{D}^{R}(t,t^{\prime}),
\ee
and to represent the time-dependent perturbation of the Hamiltonian in terms of its Fourier expansion  
\be \label{fou}
\hat{M}^{(1)}(t)=\sum_{k=-K,k \neq 0}^K\hat{M}_{k}^{(1)}e^{-ik\Omega_{0}t}.
\ee
 Substituting
 in the Dyson's equation  (\ref{dyson3}) results
\be
\label{dyson4}
\hat{D}^{R}(t,\omega)=\hat{D}^{(0)}(\omega)+\sum_{k\neq 0,-\infty}^{+\infty}e^{-ik\Omega_{0}t}\hat{D}^{R}(t,\omega+k\Omega_{0})\hat{M}_{k}^{(1)}\hat{D}^{(0)}(\omega),
\ee
where
\be
\label{d0}
\hat{D}^{(0)}(\omega)=\left[\omega^{2}\hat{I}-\hat{M}^{(0)}-\Sigma^{R}(\omega) \right]^{-1},
\ee
corresponds to  the stationary component of the retarded Green's function of the central chain connected to the reservoirs but without the effect of the time-dependent perturbations. 
Notice that because of the periodic structure of the time dependent part of the Hamiltonian, the retarded Green's function
is also periodic in time with period $\tau$.
For this reason it is useful and natural to represent this function in terms of a Fourier series as follows
\be
\label{floquet}
\hat{D}^{R}(t,\omega)=\sum_{k=-\infty}^{+\infty}e^{-ik\Omega_{0}t} \mathcal{D}(k,\omega)
\ee
where the functions $\mathcal{D}(k,\omega)$ are also known as Floquet components. 
The exact solution for the set of coupled linear equations Eq.(\ref{dyson4}) can be
obtained numerically by following the procedure introduced in
Ref. \cite{liligreen} for fermionic systems driven by ac
potentials. Alternatively, it is possible to solve these equations approximately in the limits of 
weak amplitudes $\hat{M}_k^{(1)}$ of the time-dependent perturbations and in the limit of low frequencies $\Omega_0$.
We briefly present these approximate treatments in the next subsections.
%
%%%%%%%%%%%%%%%%%%%%%%%%%%%%%%%%%%%%%%%%%%%%%%%%%%%%%%%%%%%%%%%%%%%%%%%%%%%%%%%%
\subsection{Perturbative solution of Dyson's equation}
\label{pert}
In most cases, the set of linear equations of Eq.(\ref{dyson4}) must be solved numerically. 
For this reason, it is convenient to carry out a systematic expansion in powers of
$\hat{M}^{(1)}$, to obtain analytical expressions.
When these amplitudes are small (compared with the energies of the Hamiltonian
independent of time), a perturbative solution in these parameters is
a good approximation.
Assuming that the external perturbations $\hat{M}^{(1)}(t)$ contain $K$ harmonics, as expressed in Eq.  (\ref{fou}), the Green's function evaluated up to the 1st order in
the amplitudes $\hat{M}_k^{(1)}$ is
\be \label{dp1}
\hat{D}^{R}(t,\omega)\sim \sum_{k=-K}^{K}e^{-ik\Omega_{0}t} \hat{\mathcal{D}}(k,\omega),
\ee
with
\begin{eqnarray} \label{dp2}
&& \hat{\mathcal{D}}(0,\omega)=\hat{D}^{(0)}(\omega),\\
&& \hat{\mathcal{D}}(\pm k,\omega)=\hat{D}^{(0)} (\omega \pm k \Omega_{0}) \hat{M}_{\pm k}^{(1)}\hat{D}^{(0)}(\omega),\;\;\;\; k=1,\ldots, K.
\end{eqnarray}
%
%%%%%%%%%%%%%%%%%%%%%%%%%%%%%%%%%%%%%%%%%%%%%%%%%%%%%%%%%%%%%%%%%%%%%%%%%%%%%%%%
\subsection{Low driving frequency solution of Dyson's equation}
\label{adia}
Let us now consider the Dyson equation in Eq. (\ref{dyson4}) in the limit
of low driving frequency $\Omega_0$.
%
%%%%%%%%%%%%%%%%%%%%%%%%%%%%%%%%%%%%%%%%%%%%%
\subsubsection*{1st order}
A solution exact up to ${\cal O}(\Omega_0)$ can be obtained by
expanding Eq. (\ref{dyson4}) as follows:
\ba \hat{D}(t,\omega) & \sim &\hat{D}^{(0)}(\omega) +
\hat{D}(t,\omega) \hat{M}^{(1)}(t) \hat{D}^{(0)}(\omega) + \nonumber
\\ & & i \partial_{\omega} \hat{D}(t,\omega) \frac{d
  \hat{M}^{(1)}(t)}{dt} \hat{D}^{(0)}(\omega).  \ea
We define the frozen Green function
\be \label{froz} \hat{D}_f (t,\omega) = \left[
  \hat{D}^{(0)}(\omega)^{-1} - \hat{M}^{(1)}(t) \right]^{-1}, \ee
in terms of which the exact solution of the Dyson equation at ${\cal
  O}(\Omega_0)$ reads
\be \hat{D}^{(1)}(t,\omega) = \hat{D}_f (t,\omega) + i
\partial_{\omega} \hat{D}_f (t,\omega) \frac{d \hat{M}^{(1)}(t)}{dt}
\hat{D}^{(0)}(\omega).  \ee
%
%%%%%%%%%%%%%%%%%%%%%%%%%%%%%%%%%%%%%%%%%%%%%%%%%%%%%
\subsubsection*{2nd order}
To obtain the solution exact up to ${\cal O}(\Omega_0^2)$, we consider
the following expansion of equation Eq. (\ref{dyson4}):
%
%\begin{widetext}
\ba \hat{D}(t,\omega) & \sim & \hat{D}^{(0)}(\omega) +
\hat{D}(t,\omega) \hat{M}^{(1)}(t) \hat{D}^{(0)}(\omega) \nonumber
\\ & & +\ i \partial_{\omega} \hat{D}(t,\omega) \frac{d
  \hat{M}^{(1)}(t)}{dt} \hat{D}^{(0)}(\omega) \nonumber \\ & &
-\ \frac{1}{2} \partial^2_{\omega} \hat{D}(t,\omega) \frac{d^2
  \hat{M}^{(1)}(t)}{dt^2} \hat{D}^{(0)}(\omega).  \ea
%\end{widetext}
%
The solution exact up to ${\cal O}(\Omega_0^2)$ is
\ba
\label{d2}
\hat{D}^{(2)}(t,\omega) & = & \hat{D}_f(\omega) + i \partial_{\omega}
\hat{D}^{(1)}(t,\omega) \frac{d \hat{M}^{(1)}(t)}{dt}
\hat{D}_f(\omega) \nonumber \\ & & -\frac{1}{2} \partial^2_{\omega}
\hat{D}_f (t,\omega) \frac{d^2 \hat{M}^{(1)}(t)}{dt^2}
\hat{D}_f(\omega) .  \ea
One can obtain this Green function numerically by first discretizing
the time in the interval $0 \leq t \leq \tau$, then solving
Eq. (\ref{d2}) for each one time, and finally evaluating the Fourier
transform to obtain the Floquet representation with which the dc
component of the heat current can be calculated from
Eq. (\ref{dcalph}).
%
%%%%%%%%%%%%%%%%%%%%%%%%%%%%%%%%%%%%%%%%%%%%%%%%%%%%%%%%%%%%%%%%%%%%%%%%%%%%
\section{dc heat current}
In this section, we present the steps to calculate the net heat current flowing into a given reservoir.
We start by writing  the time-dependent heat current given in Eq.(\ref{curalft}) as follows
\begin{eqnarray}
J_{\alpha}^{Q}(t)&=&\sum_{n}k_{\alpha}\sqrt{\frac{2}{N_{\alpha}+1}} \sin(q_{n}^{\alpha})\lim_{t\rightarrow t^{\prime}}\frac{\partial}{\partial t^{\prime}} \langle x_{\alpha,n}(t^{\prime})x_{l_{\alpha}}(t) \rangle\\
&=&- \sum_{n}\frac{k_{\alpha}}{m_c}\sqrt{\frac{2}{N_{\alpha}+1}} \sin(q_{n}^{\alpha})\lim_{t\rightarrow t^{\prime}}\mbox{Re}
\{i \frac{\partial}{\partial t^{\prime}} D_{l_{\alpha},q_{n}^{\alpha}}^{<}(t,t^{\prime})\}.
\end{eqnarray}
We now substitute Eq.(\ref{dyless1}) and the representation of Eq.(\ref{floquet}) to write the dc component of the above current as
\ba
\overline{J}_{\alpha}^{Q}(t) &= &- \sum_{n}k_{n,\alpha}\int_{-\infty}^{+\infty} \frac{d\omega}{2\pi}
\mbox{Re}\{ \omega \mathcal{D}_{l_{\alpha},l_{\alpha}}(0,\omega)d_{q_{n}^{\alpha}}^{<}(\omega)+ \nonumber \\
& & \sum_{\beta=1}^{L}\sum_{k=-\infty}^{+\infty}(\omega+k\Omega_0)|\mathcal{D}_{l_{\alpha},l_{\beta}}(k,\omega)|^{2}\Sigma_{\beta}^{<}(\omega)d_{q_{n}^{\alpha}}^{A}(\omega+k\Omega_{0})\}
\ea
where $k_{n,\alpha}=\left(\frac{k_{\alpha}^{2} }{m_{\alpha} m_c}\right)\frac{2}{N_{\alpha}+1} \sin^{2}(q_{n}^{\alpha})$.
Finally we obtain
\ba \overline{J}^Q_{\alpha} & = & \int_{-\infty}^{+\infty} \frac{d \omega}{2 \pi} 
\left[\omega 2 \mbox{Im}\left\lbrace{\cal D}_{l_{\alpha},l_{\alpha}}(0,\omega)\right\rbrace
\Gamma_{\alpha}(\omega) n_{\alpha}(\omega) \nonumber\right. \\ & &
+ \sum_{\beta=1}^{L} \sum_{k=-\infty}^{+\infty} (\omega+k\Omega_0)|{\cal D}_{l_{\alpha},l_{\beta}}(k,\omega)|^2
\Gamma_{\beta}(\omega)\Gamma_{\alpha}(\omega + k \Omega_0)
n_{\beta}(\omega) ]. \nonumber \\ \ea
An alternative representation to this current can be obtained after recasting the first term of the above equation by recourse to the identity
\ba
\label{id1}
 {\cal D}_{l,l^{\prime}}(k,\omega)-{\cal D}_{l^{\prime},l}^*(-k,\omega+k \Omega_0 ) & = &
-i \sum_{k^{\prime}}\sum_{\alpha} {\cal
  D}_{l,l_{\alpha}}(k+k^{\prime},\omega-k^{\prime} \Omega_0) \nonumber \\
  & & \times
\Gamma_{\alpha}(\omega-k^{\prime} \Omega_0) {\cal
  D}^*_{l^{\prime},l_{\alpha}}(k^{\prime},\omega-k^{\prime} \Omega_0).
\ea

This leads to the following equation for the dc heat current,
\be
\label{dcalph}
\overline{J}^{Q}_{\alpha} = 
\sum_{\beta=1}^{L}\sum_{k=-\infty}^{+\infty} \int_{-\infty}^{+\infty}
\frac{d \omega}{2 \pi} (\omega + k \Omega_0) [n_{\beta}(\omega)
  -\ n_{\alpha}(\omega+k\Omega_0)]
\Gamma_{\alpha}(\omega+k\Omega_0) \Gamma_{\beta}(\omega) |{\cal
  D}_{l_{\alpha},l_{\beta}}(k,\omega)|^2.
\ee
This equation constitutes a generalization to the case of ac driving of a Landauer-B\"uttiker formula for the heat transport in phononic systems (see Ref. \cite{wang})
\be
J_{\alpha}^{Q}=\int_{-\infty}^{+\infty} \frac{d\omega}{2\pi} \omega \sum_{\beta=1}^{L} {\cal T}_{\alpha \beta}(\omega)[n_{\beta}(\omega)-n_{\alpha}(\omega)],
\ee
being $ {\cal T}_{\alpha \beta}(\omega)=
|D_{l_{\alpha},l_{\beta}}^{R}(\omega)|^{2}\Gamma_{\alpha}(\omega)\Gamma_{\beta}(\omega)$, the transmission function between the reservoirs $\alpha$ and $\beta$.
Equation (\ref{dcalph}) indicates that a net heat current may exist even in the
absence of a temperature difference between the reservoirs. Such a
flow is a consequence of the pumping with the ac forces and contains, in general, an incoming component which accounts for the
work done by the ac fields and which is dissipated into the
reservoirs. As shown in Ref. \cite{liliphon} under certain conditions, a net
current between the reservoirs can be established. This current can go
against a temperature gradient, therefore allowing for cooling.
%

% % 
%%%%%%%%%%%%%%%%%%%%%%%%%%%%%%%%%%%%%%%%%%%%%%%%%%%%%%%%%%%%%%%%%%%%%%%%%%%%%%%%
\section{Mean power developed by the external forces}
We now turn to analyze another interesting aspect of the heat transport, which are the mechanisms used by the driving forces  to transfer a net amount of energy
to the phononic system. For sake of simplicity we restrict our analysis to the case of driving forces that oscillate with a single harmonic component and a phase-lag, of the form
\be \label{kl}
k_{l}^{\prime}(t)=K_0 \cos(\Omega_{0}t+\delta_{l}).
\ee

The mean power developed by the force acting on the site $l$ corresponds to calculating the dc component of the time-dependent power of 
Eq.(\ref{powla}). In terms of Green's functions can be written as follows
\be
\overline{P}_{l}=\frac{-i}{\tau m_c}\int_{0}^{\tau }dt \frac{\partial k_{l}^{\prime} (t)}{\partial t} D^<_{l,l}(t,t).
%\int dt_1\int dt_2 D^{R}_{l,l_{\alpha}}(t,t_1)\Sigma_{\alpha}^{<}(t_1-t_2)D_{l,l_{\alpha}}^{R*}(t,t_2).
%\frac{-i}{\tau}\int_{0}^{\tau}dt \frac{\partial k_{l l^{\prime}} (t)}{\partial t} \sum_{\alpha=1}^{L}
%\int dt_1\int dt_2 D^{R}_{l,l_{\alpha}}(t,t_1)\Sigma_{\alpha}^{<}(t_1-t_2)D_{l,l_{\alpha}}^{R*}(t,t_2).
\ee
According to the Dyson's equation (\ref{dyless}) and in terms of the Floquet-Fourier representation of the Green's function (\ref{floquet})
the above equation reads
\begin{eqnarray}
\overline{P}_{l}=-\frac{\Omega_{0}}{m_c} \sum_{\alpha}\sum_{k,k^{\prime}}\int_{-\infty}^{+\infty} \frac{d\omega}{2\pi} M_{k^{\prime}}^{(1)}k^{\prime}\mathcal{D}_{l,l_{\alpha}}(k,\omega)
\Sigma_{\alpha}^{<}(\omega)\mathcal{D}_{l,l_{\alpha}}^{*}(k+k^{\prime},\omega),
\end{eqnarray}
%
%where $M_{k^{\prime}}^{(1)}=\int dt e^{k^{\prime}\Omega_{0}t}k_{i}^{\prime}(t)$.
For the case of ac forces of the form (\ref{kl}) this equation results
\begin{eqnarray}
\overline{P}_{l}&=& \frac{\Omega_{0}K_0}{2 m_c}\sum_{k}\sum_{\alpha=1}^{L}\int_{-\infty}^{+\infty} \frac{d\omega}{2\pi} i n_{\alpha}(\omega)[\mathcal{D}_{l,l_{\alpha}}(k,\omega)\Gamma_{\alpha}(\omega)\mathcal{D}_{l,l_{\alpha}}^{*}(k-1,\omega)e^{i\delta_{l}}\\
&&- \mathcal{D}_{l,l_{\alpha}}(k,\omega)\Gamma_{\alpha}(\omega)\mathcal{D}_{l,l_\alpha}^{*}(k+1,\omega)e^{-i\delta_l}].
\end{eqnarray}
A simple analytical result can be derived for weak driving, in which case we can use the perturbation treatment of Sec. \ref{pert}. This leads to a result for the mean power which is exact
up to $\mathcal{O}(K_{0}^{2})$  and reads 
\begin{eqnarray}
\overline{P}_{l}&\sim & \frac{\Omega_{0}K_0^{2}}{2\pi m_c}\sum_{j=1}^{N}\int_{-\infty}^{+\infty} d\omega \mbox{Im} \left\lbrace \left[\cos(\delta_{l}-\delta_{j})-i\sin(\delta_{l}-\delta_{j})\right]\right.\\
&&\left.\left[\gamma_{l,j}(\omega) D_{l,j}^{(0)}(\omega-\Omega_{0})
 + \gamma_{l,j}^{*}(\omega) D_{l,j}^{(0)*}(\omega+\Omega_{0})\right]\right\rbrace,
\end{eqnarray}
%\mathcal{D}_{i,i_{\alpha}}^{(0)}(\omega)\Gamma_{\alpha}(\omega)\mathcal{D}_{j,i_{\alpha}}^{(0)*}(\omega)
with 
\be
\gamma_{l,j}(\omega)=\sum_{\alpha=1}^{L}n_{\alpha}(\omega){D}_{l,l_{\alpha}}^{(0)*}(\omega)\Gamma_{\alpha}(\omega)\mathcal{D}_{j,l_{\alpha}}^{(0)}(\omega).
\ee
It is particularly interesting to analyze this result in the limit of low driving frequency  $\Omega_{0}$. To this end we expand the previous equation up to second order in this parameter. In the ensuing
result we find that in this limit the mean power contains two different kinds of components $\overline{P}_l= \overline{P}_l^{ex}+\overline{P}_l^{dis}$, being
\begin{eqnarray}
\overline{P}^{ex}_{l}& = & -\frac{\Omega_{0}K_0^{2}}{\pi m_c}\sum_{j=1}^{N} \sin(\delta_{l}-\delta_{j}) \int_{-\infty}^{+\infty} d\omega
\mbox{Re}\left\lbrace \gamma_{l,j}(\omega) D_{l,j}^{(0)}(\omega)\right\rbrace, \\
\overline{P}^{dis}_{l}&= & - \frac{\Omega_{0}^2 K_0^{2}}{\pi m_c}\sum_{j=1}^{N}
\cos(\delta_{l}-\delta_{j}) \int_{-\infty}^{+\infty} d\omega \mbox{Im}\left\lbrace \gamma_{l,j}(\omega)\frac{\partial D_{l,j}^{(0)} (\omega)}{\partial \omega}\right\rbrace,
\end{eqnarray}
where the first component is $\propto \Omega_0$ and it is the leading contribution for low driving frequency. Since $\sin(\delta_l-\delta_j)$ is an antisymmetric function, this term satisfies 
\be
\sum_{l=1}^N \overline{P}^{ex}_l =0,
\ee
indicating that some of the forces deliver, while others receive a net amount of energy. This is the essential ingredient for this setup to have an operational regime typical of an engine. In the present case, it is particularly appealing The fact that the exchanged energy is of mechanical nature. The second
 component, which is $ \propto \mathcal{O}(\Omega_0^{2})$, is non-vanishing and accounts for the conservation of the energy expressed in Eq. (\ref{cons}). This component, thus, describes the net
 amount of energy that is dissipated into the reservoirs in the form of heat. It is important to notice that  for a driving like (\ref{kl}), the exchange component is non-vanishing only in the case of a driving with a phase-lags 
 $\delta_l-\delta_j \neq n \pi$.  Another possibility to get finite values for $\overline{P}^{ex}_{l}$ is to consider local forces with different amplitudes. It can be in general shown that a multiple-parametric
 driving is a necessary condition to have the exchange mechanism in the present setup.

 \section{Summary and conclussions}
 We have considered a simple model to study heat transport in nanomechanical systems. It consists in the usual model for acoustic phonons coupled to phononic reservoirs and generalized to include
 the effect of a time-dependent perturbation in the form of elastic forces acting at different places of the structure.  We have reviewed a theoretical treatment based on non-equilibrium Green's 
 functions, which is analogous to the one previously proposed for electron systems under ac-driving. We have derived the equations for the heat current along different regions of the setup and shown how can
 be evaluated in terms of the Green's functions. We have also derived expressions to evaluate the power developed by the external ac forces.  In the limit of weak driving amplitudes and low frequencies,
 we have analytically shown that the mean power done by each force, contains a component that represents the heat dissipated into the reservoirs. For a driving with multiple parameters, corresponding for instance to driving forces oscillating with a phase lag, there is an additional component, which corresponds to energy
 that can be exchanged between the different forces. This component dominates for low enough driving frequencies  and enables an operational regime of the setup as a quantum engine. 
  A similar mechanism has been identified in driven electron systems like  quantum pumps \cite{liliheat} and in coupled quantum capacitors
 \cite{mobu}. The present case has the appealing feature of involving mechanical work. 
 
 The present theoretical approach sets a general framework for the study of heat transport by vibrational degrees of freedom in driven systems. It can be adapted, generalized and improved in many directions,
 to analyze several interesting situations. In particular, we have considered phononic reservoirs, but the present treatment can be easily adapted to deal with photonic reservoirs. We have considered coupled atoms, but this treatment can be easily generalized to consider coupled molecules, which have internal vibrational degrees of freedom. Improvements to model realistic setups with many vibrational modes can also be straightforwardly implemented. Extensions of the present treatment to consider the anharmonic effects of Ref. \cite{wang} with ac driving are also possible. To finalize, it is  interesting to mention that  a family of 1d interacting models of cold atoms can be reduced to a Luttinger Hamiltonian, which is basically a model of harmonic oscillators. Dynamical evolution of such systems in quenching protocols including coupling to reservoirs have been recently considered \cite{leto}. The present scheme could be useful to study such systems under ac driving. 
 
\section{Acknowledgments}
LA thanks R. Capaz, C. Chamon and E. Mucciolo for many discussions.
We acknowledge  support from CONICET, UBACyT and MINCYT through PICT, Argentina.  
 
%%%%%%%%%%%%%%%%%%%%%%%%%%%%%%%%%%%%%%%%%%%%%%%%%%%%%%%%%%%%%%%%%%%%%%%%%%%%
\appendix
\section{Self-energy corresponding to the coupling to a semi-infinite chain of acoustic phonons} \label{apa}
The explicit evaluation of the sum over the normal modes entering the expression of the self energy (\ref{sigal}) can be expressed in terms of a spectral density $\Gamma_{\alpha}(\omega)$ as follows
\be
\Sigma_{\alpha}^{R}(t,t^{\prime})=\int_{-\infty}^{+\infty}\frac{d\omega}{2\pi} e^{-i\omega(t-t^{\prime})}\int_{-\infty}^{+\infty}\frac{d\omega^{\prime}}{2\pi}\frac{\Gamma_{\alpha}(\omega^{\prime})}{\omega-\omega^{\prime}+i\eta},
\ee
with $\eta >0$ and
\begin{eqnarray}
\Gamma_{\alpha}(\omega)&=& \lim_{N_{\alpha} \rightarrow \infty} \frac{
  2 \pi k_{ \alpha }^2 }{(N_{\alpha}+1)m_{c}m_\alpha}
\sum_{n=0}^{N_{\alpha}}\sin^2(q^{\alpha}_n) \frac{1}{E_{\alpha,n}}
\nonumber \\ & & \times \left[ \delta(\omega-E_{\alpha,n}) -
  \delta(\omega+E_{\alpha,n}) \right] \nonumber \\ &= &
\mbox{sgn}(\omega) \frac{k_{\alpha} }{m_c} \Theta \Big( 1- \big(
\frac{ k_{\alpha}-m_{\alpha} \omega^2}{ k_{\alpha} } \big)^2 \Big)
\nonumber \\ & & \times \sqrt{ 1- \left(\frac{ k_{\alpha}-m_{\alpha}
    \omega^2}{ k_{\alpha} } \right)^2},
\end{eqnarray}
being  $E_{\alpha,n} =
\sqrt{k_{\alpha}(1 - \cos q^{\alpha}_{n} )/m_{\alpha} }$. 
Notice that the Fourier transform of the self energy is perfectly define because is 
an equilibrium function.

\end{document}